# Independent Manipulation of Heat and Electrical Current via Bifunctional Metamaterials

Massimo Moccia[1], Giuseppe Castaldi[1], Salvatore Savo[2], Yuki Sato[2,*], Vincenzo Galdi[1,*]

(1) Waves Group, Department of Engineering, University of Sannio, I-82100, Benevento, Italy
(2) Rowland Institute at Harvard, Harvard University, Cambridge, MA 02142, USA

*Corresponding authors: sato@rowland.harvard.edu, vgaldi@unisannio.it

## Abstract

Spatial tailoring of the material constitutive properties is a well-known strategy to mold the local flow of given observables in different physical domains. Coordinate-transformation-based methods (e.g., transformation optics) offer a powerful and systematic approach to design anisotropic, spatially-inhomogeneous artificial materials ("metamaterials") capable of precisely manipulating wave-based (electromagnetic, acoustic, elastic) as well as diffusion-based (heat) phenomena in a desired fashion. However versatile these approaches have been, most designs have so far been limited to serving *single-target* functionalities in a given physical domain. Here we present a step towards a "transformation multiphysics" framework that allows independent and simultaneous manipulation of multiple physical phenomena. As a proof of principle of this new scheme, we design and synthesize (in terms of realistic material constituents) a metamaterial shell that simultaneously behaves as a thermal concentrator and an electrical "invisibility cloak". Our numerical results open up intriguing possibilities in the largely unexplored phase space of multi-functional metadevices, with a wide variety of potential applications to electrical, magnetic, acoustic, and thermal scenarios.

Subject Areas: Metamaterials, Materials Science, Interdisciplinary Physics.

## I. INTRODUCTION

Traditionally, conventional materials have been devised and engineered to serve only *single-target* applications. In an integrated circuit, for example, each component is designed to play a specific role: metallic interconnection lines carry electric currents, while a separated block works as a heat sink for dissipating heat. If a single building block could be designed to perform multiple functions in different physical domains, independently but at the same time, this could lead to a completely new way to design complex systems. Natural media are not conceived to accomplish multiple functionalities at the same time, and for this reason taming different physical phenomena at will is a tough proposition.

A new avenue could be paved with the employment of properly engineered artificial materials. Driven by the ability to induce physical responses absent in Nature, the field of "metamaterials" has seen a tremendous growth in recent years. One of the catalysts for the progress made in this field, theoretically as well as experimentally, has been the so-called "transformation optics" theory [1,2]. Viewing the rerouting of energy flow as a distortion of space from a coordinate transformation, the correspondence between constitutive material parameters and geometric transformations can serve as a powerful recipe for designing and fabricating artificial structures. This approach has been utilized not only for the manipulation of electromagnetic waves [3-5], but also for acoustics [6-10], elastodynamics [11-16], electrostatic [17-19] and magnetostatic [20-24] fields, as well as liquid surface waves [25] and diffusive heat flow [26-28]. The reader is referred to [29-31] for recent perspectives and reviews of metamaterial applications to diverse fields. Within this framework, also worth of mention are some recent multiphysics studies aimed at exciting surface plasmon polaritons in graphene via the interplay of light and sound waves [32,33].

From the mounting experimental applications in various physical branches, it is clear that the strength of the transformation-optics theory is first and foremost its unconventional versatility. Taking advantage of it, one may envision applying the theory to simultaneously manipulate multiple physical phenomena in independent fashions. For example, a material may be designed to exhibit a particular thermal functionality while its electrical functionality is made drastically different via separate but intertwined coordinate transformations.

Through the example of designing a metamaterial shell that behaves as a thermal concentrator and an electrical "invisibility cloak" at the same time, we present here a framework that allows access to the phase space of multi-functionality with metastructures. Utilizing coordinate transformations while effectively linking phenomena in multiple physical domains, we demonstrate a step towards a general platform that can be called the "transformation multiphysics".

Accordingly, the rest of the paper is organized as follows. In Sec. II, with specific reference to the thermal and electrical scenarios, we outline the modeling aspects pertaining to the transformation media, their effective-medium implementation, and the numerical simulations (with details relegated in Appendices A and B). In Sec. III, we illustrate a proof-of-principle example of synthesis in terms of realistic material constituents. In Sec. IV, we provide further insight into the response exhibited by our metastructure, as well as some bounds dictated by practical limitations. Finally, in Sec. V, we provide some brief conclusions and perspectives.

## II. MODELING ASPECTS
### A. Thermal and Electrical Transformation Media

Although, in principle, our approach could be applied to different physical domains, our focus here is on the thermal and electrical responses. As illustrated in Fig. 1(a), we begin by considering an auxiliary space $r' = (x', y', z')$, filled with an isotropic medium of thermal and electrical conductivities $\kappa'$ and $\sigma'$. At equilibrium, the stationary heat and electrical conduction equations in the absence of sources are given by

$$\nabla \cdot (\kappa' \nabla T') = 0, \quad \nabla \cdot (\sigma' \nabla V') = 0, \tag{1}$$

with $T'$ and $V'$ denoting the temperature and electrical potential, respectively. In the homogeneous case (i.e., $\kappa'$ and $\sigma'$ constant), if temperature and potential differences exist at the two boundaries, the heat-flux and electrical current-density would be directed along straight, parallel paths, as schematically depicted in Fig. 1(a). This is the typical behavior of natural materials.

Next, we introduce two coordinate transformations to a new *curved-coordinate* space $r$, namely,

$$r' = \begin{cases} F_t(r), \\ F_e(r), \end{cases} \quad (2)$$

with the subscripts "$t$" and "$e$" denoting the thermal and electrical domains, which induce different local metric distortions in the two physical domains. For instance, as shown in Fig. 1(a), we consider a concentrator-type transformation in the thermal domain, and an invisibility-cloak-type transformation in the electrical domain. By exploiting the form-invariance properties of Eqs. (1), the temperature and potential distributions in the transformed domains can be readily related to the original quantities as [31]:

$$T(r) = T'[F_t(r)], \quad V(r) = V'[F_e(r)]. \quad (3)$$

Moreover, the distortion effects induced by the coordinate transformations can be equivalently obtained in a *flat*, Cartesian space $r = (x, y, z)$ filled with an inhomogeneous, anisotropic "transformation medium" [cf. Fig. 1(b)] characterized by thermal and electrical conductivity tensors [31]

$$\vec{\kappa} = \kappa' \det(\vec{\Lambda}_t) \vec{\Lambda}_t^{-1} \cdot \vec{\Lambda}_t^{-T}, \quad \vec{\sigma} = \sigma' \det(\vec{\Lambda}_e) \vec{\Lambda}_e^{-1} \cdot \vec{\Lambda}_e^{-T}, \quad (4)$$

with

$$\vec{\Lambda}_t = \frac{\partial F_t}{\partial r}, \quad \vec{\Lambda}_e = \frac{\partial F_e}{\partial r} \quad (5)$$

denoting the Jacobian matrices associated with the two coordinate transformations, and the superscripts "$-1$" and "$-T$" indicating the inverse and inverse-transpose, respectively. In such a medium, the heat-flux and electrical current-density would follow *markedly different* paths. For instance, in the concentrator/cloak example chosen, the heat-flux would tend to concentrate in the inner region, whereas the current-density would tend to circumvent that region, as schematically depicted in Fig. 1(b).

### B. Effective-Medium Modeling and Synthesis

Although it is generally impossible to find a natural material exhibiting the desired constitutive relationships in Eqs. (4), these can be approximated to a certain extent by means of metamaterials. Results available in the literature [18,27,28] deal with the design of a single functionality (e.g., cloak or concentrator) in a single domain (e.g., thermal or electrical), and the only example of bifunctional device implements the same functionality in both thermal and electrical domains [34]. Here, the task requires us to prescribe different functionalities in multiple domains, and we proceed by following a synthesis approach based on the mixture of $N$ different types of material inclusions embedded in a host medium [Fig. 1(c)]. The host and inclusions are characterized by their thermal and electrical conductivities $\kappa_n$ and $\sigma_n$, respectively, and filling fractions $f_n$, with $n = 0,1,\ldots,N$, with the subscript "0" denoting the host medium. Each inclusion is also characterized by a depolarization tensor $\vec{\vec{\Gamma}}_n$, which depends on its shape and orientation [35]. We are therefore led to finding the material and structural compound parameters $\boldsymbol{\kappa} = \{\kappa_0, \kappa_1, \ldots, \kappa_N\}$, $\boldsymbol{\sigma} = \{\sigma_0, \sigma_1, \ldots, \sigma_N\}$, $\boldsymbol{f} = \{f_0, f_1, \ldots, f_N\}$, $\vec{\vec{\Gamma}} = \{\vec{\vec{\Gamma}}_1, \ldots, \vec{\vec{\Gamma}}_N\}$ so that

$$\begin{cases} \vec{\vec{\kappa}}_{eff}\left(\boldsymbol{\kappa}, \boldsymbol{f}, \vec{\vec{\Gamma}}\right) = \vec{\vec{\kappa}}_{nom}, \\ \vec{\vec{\sigma}}_{eff}\left(\boldsymbol{\sigma}, \boldsymbol{f}, \vec{\vec{\Gamma}}\right) = \vec{\vec{\sigma}}_{nom}, \end{cases} \quad (6)$$

where $\vec{\vec{\kappa}}_{nom}$ and $\vec{\vec{\sigma}}_{nom}$ are the desired nominal constitutive tensors [arising from Eqs. (4)], whereas $\vec{\vec{\kappa}}_{eff}$ and $\vec{\vec{\sigma}}_{eff}$ are the effective constitutive tensors characterizing the mixture, which can be related to the host and inclusion parameters via approximate mixing formulae [35] (see also Appendix A for details). We highlight the *nonlinear* character of Eqs. (6) (stemming from the mixing formulae), and the *coupling* between the thermal and electrical domains via the structural compound parameters $\boldsymbol{f}$ and $\vec{\vec{\Gamma}}$. Moreover, the search space is constrained by the passivity requirements $\kappa_n \geq 0$ and $\sigma_n \geq 0$, as well as by the self-consistency conditions $0 < f_n < 1$, $\sum_{n=0}^{N} f_n = 1$, and unit-trace conditions $\text{tr}(\vec{\vec{\Gamma}}_n) = 1$. Overall, solving Eqs. (6) represents a formidable task.

The synthesis is significantly simplified if the *same* functionality is required in both domains, as in [34]. In this case, $\boldsymbol{F}_t = \boldsymbol{F}_e$ and [from Eqs. (4)] $\vec{\vec{\kappa}}_{nom}/\kappa' = \vec{\vec{\sigma}}_{nom}/\sigma'$, which implies that the problems in Eqs. (6) are *decoupled*, with only one synthesis needed. Here, the scenario is more complex, and to induce two *distinct* functionalities we exploit the concept of "neutral" inclusions from the theory of composites [36,37], i.e., inclusions that are matched with the host medium in one physical domain, so that they are effectively "visible" only in the other domain. This assumption too decouples the thermal and electrical syntheses in Eqs. (6), but it does not constrain the two functionalities to be identical. Clearly, working with natural material constituents, the required neutrality conditions may only be fulfilled approximately. Nevertheless, in principle, such inclusions may be properly engineered via multilayered composites, e.g., along the lines of [38,39].

Focusing on a two-dimensional scenario in the associated $(\rho,\phi,z)$ cylindrical coordinate system, and letting $\kappa_{\rho,nom}$, $\kappa_{\phi,nom}$, $\sigma_{\rho,nom}$, $\sigma_{\phi,nom}$ the nominal values of relevant constitutive-tensor components to be synthesized, and $\kappa_{\rho,eff}$, $\kappa_{\phi,eff}$, $\sigma_{\rho,eff}$, $\sigma_{\phi,eff}$ the corresponding effective parameters, we consider a three-phase mixture featuring two types of elliptic cylindrical inclusions with axes locally oriented along the cylindrical coordinates $\rho$ and $\phi$. The search parameter space therefore comprises the constitutive parameters $\boldsymbol{\kappa}=\{\kappa_0,\kappa_1,\kappa_2\}$ and $\boldsymbol{\sigma}=\{\sigma_0,\sigma_1,\sigma_2\}$, filling fractions $\boldsymbol{f}=\{f_0,f_1,f_2\}$ (with $f_0+f_1+f_2=1$), and relevant depolarization-tensor components $\Gamma_{1\rho}=1-\Gamma_{1\phi}$ and $\Gamma_{2\rho}=1-\Gamma_{2\phi}$. These latter components, for an elliptical inclusion with axes $A_\rho$ (along the $\rho$ direction) and $A_\phi$ (along the $\phi$ direction), are given by [35]

$$\Gamma_\rho = 1-\Gamma_\phi = \frac{A_\phi}{A_\rho + A_\phi}. \tag{7}$$

Assuming that the type-1 inclusions are *thermally neutral* $(\kappa_1=\kappa_0)$ and the type-2 inclusions are *electrically neutral* $(\sigma_2=\sigma_0)$, and considering standard Maxwell-Garnett mixing formulae [35], the effective parameters can be written as follows (see Appendix A for details):

$$\frac{\kappa_{\rho,eff}}{\kappa_0} = \frac{\kappa_0+(\kappa_2-\kappa_0)\left[\Gamma_{2\rho}(1-f_2)+f_2\right]}{\kappa_0+\Gamma_{2\rho}(1-f_2)(\kappa_2-\kappa_0)}, \tag{8}$$

$$\frac{\kappa_{\phi,eff}}{\kappa_0} = \frac{\kappa_0+(\kappa_2-\kappa_0)\left[\Gamma_{2\phi}(1-f_2)+f_2\right]}{\kappa_0+\Gamma_{2\phi}(1-f_2)(\kappa_2-\kappa_0)}, \tag{9}$$

$$\frac{\sigma_{\rho,eff}}{\sigma_0} = \frac{\sigma_0+(\sigma_1-\sigma_0)\left[\Gamma_{1\rho}(1-f_1)+f_1\right]}{\sigma_0+\Gamma_{1\rho}(1-f_1)(\sigma_1-\sigma_0)}, \tag{10}$$

$$\frac{\sigma_{\phi,eff}}{\sigma_0} = \frac{\sigma_0+(\sigma_1-\sigma_0)\left[\Gamma_{1\phi}(1-f_1)+f_1\right]}{\sigma_0+\Gamma_{1\phi}(1-f_1)(\sigma_1-\sigma_0)}. \tag{11}$$

By substituting Eqs. (8)-(11) in Eqs. (6), we observe that the synthesis problem is now *effectively decoupled*, as the thermal parameters in Eqs. (8) and (9) do not depend any longer on the type-1 inclusions, whereas the electrical parameters in Eqs. (10) and (11) do not depend on the type-2 inclusions. Under these conditions, the synthesis problem can be solved analytically in closed form. Referring to Appendix A for the general solution, we consider here the limiting case $\kappa_2 \ll \kappa_0$ and $\sigma_1 \ll \sigma_0$, which yields the particularly simple results

$$\kappa_1 = \kappa_0 = \kappa_{\phi,nom} \frac{1-\Gamma_{2\phi}(1-f_2)}{(1-\Gamma_{2\phi})(1-f_2)}, \quad \Gamma_{2\phi} = \frac{1-\bar{\kappa}_{nom}(1-2f_2)-\sqrt{(1-\bar{\kappa}_{nom})^2 + 4f_2^2 \bar{\kappa}_{nom}}}{2(1-f_2)(1-\bar{\kappa}_{nom})}, \quad (12)$$

$$\sigma_2 = \sigma_0 = \sigma_{\phi,nom} \frac{1-\Gamma_{1\phi}(1-f_1)}{(1-\Gamma_{1\phi})(1-f_1)}, \quad \Gamma_{1\phi} = \frac{1-\bar{\sigma}_{nom}(1-2f_1)-\sqrt{(1-\bar{\sigma}_{nom})^2 + 4f_1^2 \bar{\sigma}_{nom}}}{2(1-f_1)(1-\bar{\sigma}_{nom})}, \quad (13)$$

where $\bar{\sigma}_{nom} = \sigma_{\phi,nom}/\sigma_{\rho,nom}$, $\bar{\kappa}_{nom} = \kappa_{\phi,nom}/\kappa_{\rho,nom}$, and the filling fractions $f_1$ and $f_2$ appear as free parameters. It can be readily verified that the results in Eqs. (12) and (13) are *inherently feasible*, as they yield $\kappa_0 \geq 0$, $\sigma_0 \geq 0$, $0 < \Gamma_{1\phi} < 1$, and $0 < \Gamma_{2\phi} < 1$, for *arbitrary* values of the nominal anisotropy ratios $\bar{\kappa}_{nom}$ and $\bar{\sigma}_{nom}$. However, practical considerations (related to the spatial arrangement of the inclusions) as well as model-consistency issues effectively restrict the attainable anisotropy ratios to *moderate* values. These aspects are discussed in more detail in Sec. IV-B below.

### C. Numerical Modeling

All numerical simulations of the thermal and electrical responses in our study are carried out by means of COMSOL Multiphysics 4.2, a finite-element-based commercial software package that allows multiphysics simulations in the presence of anisotropic, inhomogeneous constitutive parameters [40]. In particular, for our simulations, we utilize the "Heat Transfer" and "AC/DC" modules [40] in order to solve the stationary, sourceless heat and electrical conduction equations, $\nabla \cdot (\vec{\kappa} \cdot \nabla T) = 0$ and $\nabla \cdot (\vec{\sigma} \cdot \nabla V) = 0$, in the square computational domain shown in Fig. 2. The thermal and electrical conductivities are generally described by tensor, inhomogeneous quantities ($\vec{\kappa}$ and $\vec{\sigma}$, respectively), which reduce to scalar, piecewise-homogeneous quantities in the inclusion-based implementations. For boundary conditions, we set a temperature difference $\Delta T = T_3 - T_1$ and a potential difference $\Delta V = V_3 - V_1$ between the right and left boundaries (labeled as 3 and 1, respectively, in Fig. 2), and enforce the thermal- and electrical-insulation conditions $\hat{\boldsymbol{n}}\cdot(\ddot{\bar{\kappa}}\cdot\nabla T)=0$, $\hat{\boldsymbol{n}}\cdot(\ddot{\bar{\sigma}}\cdot\nabla V)=0$, at the two remaining boundaries (labeled as 2 and 4, in Fig. 2), with $\hat{\boldsymbol{n}}$ denoting the outward normal unit-vector.

The computational domain is discretized via triangular meshing (using default criteria), resulting in a number of elements that, for the more complex inclusion-based structures, can be on the order of $10^7$, i.e., about 15 million degrees of freedom. For the solution of the discretized problem, we use the SPOOLES direct solver, with default parameters [40]. For the more complex inclusion-based structures, typical simulations (on a dedicated workstation with quad-core Intel Core i-7 3.40 GHz processor, 16 GB RAM, running 64bit Windows8) may require up to 13 hours.

The observables shown in the figures below are the total heat-flux $\ddot{\bar{\kappa}}\cdot\nabla T$ (W/m$^2$) and current-density $\ddot{\bar{\sigma}}\cdot\nabla V$ (A/m$^2$), normalized by the enforced quantities $\kappa'\Delta T/L$ and $\sigma'\Delta V/L$, respectively (with $L$ denoting the sidelength of the square computational domain; cf. Fig. 2). More specifically, the magnitudes of these (vector) observables are represented in false-color scale, while their local directions are indicated by the superimposed streamlines.

## III. PROOF-OF-PRINCIPLE EXAMPLE

### A. Thermal Concentrator and Electrical Cloak

The above synthesis procedure can be applied to the scenario illustrated in Fig. 1 by introducing two (scalar) radial coordinate transformations

$$\rho' = \begin{cases} F_t(\rho), \\ F_e(\rho), \end{cases} \tag{14}$$

for which Eqs. (4) can be particularized in terms of the relevant components [34]

$$\kappa_\rho = \frac{\kappa'^2}{\kappa_\phi} = \kappa'\frac{F_t(\rho)}{\rho \dot{F}_t(\rho)}, \quad \sigma_\rho = \frac{\sigma'^2}{\sigma_\phi} = \sigma'\frac{F_e(\rho)}{\rho \dot{F}_e(\rho)}, \tag{15}$$

with the overdot denoting differentiation with respect to the argument. As schematically illustrated in Fig. 3(a), the transformations in the thermal and electrical domains map an annular cylinder of radii $R_1 = 2cm$ and $R_2 = 12cm$ in the transformed space $r$ onto an annular cylinder of radii $R_c > R_1$ and $R_2$ and a cylinder of radius $R_2$, respectively, in the auxiliary space $r'$. From the functional viewpoint, $F_t$ yields a concentration effect (with $c = R_c/R_1 > 1$ denoting the concentration factor), whereas $F_e$ yields an invisibility cloaking effect. In order to achieve these effects, only the boundary values

$$F_e(R_1) = 0, \quad F_t(R_1) = cR_1, \quad F_t(R_2) = F_e(R_2) = R_2 \tag{16}$$

are prescribed, whereas the function behaviors in between are only partially constrained (see Appendix B for more details). In our example below, we exploited this degree of freedom by selecting the two mapping functions so that

$$\frac{\kappa_\rho}{\kappa_\phi} = \frac{\sigma_\phi}{\sigma_\rho}. \tag{17}$$

Though not strictly necessary, this choice allows us to utilize two types of inclusions with identical shape (just rotated of 90°) and filling fractions, which arguably facilitates their spatial arrangement (see Sec. III-B and Appendix B for details). Figure 3(b) shows the corresponding profiles for the constitutive parameters $\kappa_\rho = \kappa'^2/\kappa_\phi$ and $\sigma_\rho = \sigma'^2/\sigma_\phi$. We observe that an *exact* implementation of the transformations would require *extreme* parameters (either zero or infinite) at the inner boundary $\rho = R_1$. Acknowledging the aforementioned practical limitations, we approximate the continuous parameter distributions in terms of six-layer piecewise-constant profiles [indicated by dashed lines and markers in Fig. 3(b)], with truncation of the (extreme) parameters so as to limit the anisotropy ratio $\kappa_\rho/\kappa_\phi = \sigma_\phi/\sigma_\rho$ to moderate values $\leq 2.5$ (see also the discussion in Sec. IV-B below). Figures 3(c) and 3(d) show the corresponding thermal (concentrator) and electrical (cloak) responses, respectively. As it can be observed, in the exterior region $\rho > R_2$ the two responses are essentially identical with those observed in the unperturbed background medium (constant heat-flux and current-density, and straight, parallel streamlines), whereas they differ substantially inside the transformation-medium shell and in the inner region. More specifically, the thermal response [Fig. 3(c)] resembles that of a concentrator, with streamlines focusing toward the inner region, wherein an enhancement of the enforced heat-flux by a factor $1.53$ is attained. Conversely, the electrical response [Fig. 3(d)] resembles that of an (imperfect) invisibility cloak, with only little penetration of the streamlines in the inner region, wherein a reduction of the enforced current-density by a factor $0.55$ is attained.

### B. Preliminary Ideal-Parameter Metamaterial Synthesis

Starting from the six-layer nominal profiles in Fig. 3(b), for each sampled value, we extract the scaled conductivities $\kappa_\rho/\kappa'$, $\kappa_\phi/\kappa'$, $\sigma_\rho/\sigma'$ and $\sigma_\phi/\sigma'$ (with $\kappa'$ and $\sigma'$ denoting the background parameters), and compute the nominal anisotropy ratios $\bar{\kappa}_{nom}$ and $\bar{\sigma}_{nom}$. Next, we choose the filling fractions $f_1 = f_2$ taking into account the aforementioned assumptions and limitations (see also Appendices A and B). As a rule of thumb, taking into account that the transformation media to synthesize tend to become more isotropic towards the exterior layers (see Appendix B), we assume a gradually decreasing law (from interior to exterior layers) fulfilling the bound $f_1 = f_2 \leq 0.2$ (see also our discussion in Sec. IV-B below). Assuming neutral inclusions ($\kappa_1 = \kappa_0$ and $\sigma_2 = \sigma_0$) with $\kappa_2 \ll \kappa_0$ and $\sigma_1 \ll \sigma_0$ (assumed, for simplicity, $\kappa_2 = 0$ and $\sigma_1 = 0$), we now have all the entries in Eqs. (12) and (13) to compute the unknown parameters $\kappa_0/\kappa'$, $\sigma_0/\sigma'$, $\Gamma_{1\phi} = 1 - \Gamma_{1\rho}$, and $\Gamma_{2\phi} = 1 - \Gamma_{2\rho}$.

Table I shows, for each layer, the computed parameters. As anticipated (see also Appendix B), we observe that, in view of the particular choice in Eq. (17), we obtain $\Gamma_{2\phi} = 1 - \Gamma_{1\phi} = \Gamma_{2\rho}$, i.e., the two types of inclusions have identical shape (just rotated of 90°). As a consequence, from Eqs. (9) and (10), we also obtain that $\kappa_0/\kappa' = \sigma_0/\sigma'$. Assuming elliptical inclusions, also shown in Table I are the axis ratios calculated from the depolarization factors [35],

$$\frac{A_\rho}{A_\phi} = \frac{\Gamma_\phi}{1 - \Gamma_\phi} = \frac{\Gamma_\phi}{\Gamma_\rho}. \tag{18}$$

Table I provides the geometrical, structural, and constitutive parameters for the host medium and the two types of inclusions in each layer, which constitutes all the information needed for an inclusion-based implementation.

Based on this information, we generated the geometry in Fig. 4(a) [with magnified details shown in Figs. 4(b) and 4(c)], via heuristic placement of the elliptical inclusions, oriented along the local $\rho$ (and $\phi$) directions [see Fig. 4(d)], in different host materials. In particular, we generated and replicated an angular sector of aperture $\sim 3°$, by manually placing the elliptical inclusions with the prescribed axis-ratios, and with number and size chosen according to the prescribed filling fractions. We found that working with the two types of inclusions having same shape (just rotated of 90°) and filling fractions significantly facilitates their spatial arrangement. However, although particular care was taken in ensuring a uniform and spatially efficient packing, avoiding dense concentrations, the procedure is not optimized, and there is room for further improvement.

Figures 4(e) and 4(f) show the corresponding thermal and electrical responses, respectively, assuming a background medium with $\kappa' = 1\,W/(mK)$ and $\sigma' = 1\,S/m$. By comparison with the nominal-parameter predictions in Figs. 3(c) and 3(d), we observe a generally good agreement, both in the qualitative behaviors of the streamlines (which tend to focus in the inner region in the thermal case, and to circumvent it in the electrical case), and in the quantitative figures of merit. In particular, in the inner region, the enforced heat-flux is enhanced by a factor 1.51 (concentrator), while the current-density is reduced by a factor 0.55 (cloak).

### C. Realistic-Parameter Metamaterial Synthesis

The above design is idealized in the sense that assumes the availability of host materials with strictly prescribed constitutive parameters (cf. Table I) and neutral inclusions, which, in practice, may only be somehow approximated.

While maintaining the same geometrical and structural parameters as those in Table I and Figs. 4(a)-4(c), we tried to approximate this ideal-parameter configuration by utilizing only five realistic material constituents (detailed in Table II), via a heuristic matching with a list of realistic material parameters [41]. Also in this case, the procedure is not optimized, and further improvements are possible. Nevertheless, the obtained configuration provides a proof of principle of the practical feasibility of our transformation-multiphysics approach.

Figures 5(a) and 5(b) show the corresponding thermal and electrical responses, respectively, which are in good agreement with the nominal-parameter predictions [Figs. 3(c) and 3(d)] and the previous ideal-parameter synthesis [Figs. 4(e) and 4(f)], in spite of the imperfect fulfillment of the neutral-inclusion conditions. In particular, in the inner region, the enforced heat-flux is enhanced by a factor 1.45 (concentrator), while the enforced current-density is reduced by a factor 0.52 (cloak), once again in good agreement with the nominal-parameter predictions.

## IV. SOME REMARKS

### A. Comparison with Conventional Material Shell

To better understand the effects of our bifunctional metamaterial shell, it is insightful to compare its thermal and electrical responses with a reference configuration based on a conventional material. To give an idea, Figs. 6(a) and 6(b) show the thermal and electrical response, respectively, of a shell of same size made of stainless steel ($\kappa = 16.3 \, W/(mK)$, $\sigma = 1.42 \cdot 10^6 \, S/m$) [42], and immersed in the same P-55S 2k ($\kappa = 120 \, W/(mK)$, $\sigma = 1.18 \cdot 10^5 \, S/m$) background medium as in Fig. 5, with identical boundary conditions. By comparing the electrical responses in Figs. 6(b) and 5(b), we note a similar reduction (by a factor 0.58) of the current-density in the inner region. However, for the stainless-steel shell this also implies a sensible reduction (by a factor 0.3) of the heat-flux [compare Fig. 6(a) with Fig. 5(a)]. Moreover, both the thermal and electrical responses are significantly perturbed in the area surrounding the shell. Conversely, our metamaterial shell is capable of enhancing the heat-flux in the inner region, while reducing the current-density, with very weak effects in the exterior background region.

### B. Realistic Anisotropy Bounds

As anticipated in Sec. II-B, although our synthesis procedure in Eqs. (12) and (13) inherently yields physically feasible parameters, irrespective of the nominal anisotropy ratios $\bar{\kappa}_{nom}$ and $\bar{\sigma}_{nom}$, there are certain practical limitations to account for. It can be verified that extreme anisotropy ratios require extreme values of the depolarization factors and/or high values of the filling fractions. The former requirement ($\Gamma_\phi \to 0$ or $\Gamma_\phi \to 1$) translates into *needle-shaped* inclusions ($A_\rho \ll A_\phi$ or $A_\phi \ll A_\rho$) that may be difficult to arrange in a spatially efficient fashion. The latter requirement (relatively high filling fractions), on the other hand, entails significantly dense mixtures, for which the assumed Maxwell-Garnett mixing formulae may not represent an adequate model [35].

It is therefore important to estimate some realistic bounds on the anisotropy ratios that arise from these limitations, and the consequent constraints in the coordinate-transformations that may be implemented. In our design procedure above, we found that values of the depolarization factors $0.1 \leq \Gamma_{1,2\phi} \leq 0.9$ and of the filling fractions $f_{1,2} \leq 0.2$ usually allow spatially-efficient arrangements of the inclusions, which are also adequately modeled by the Maxwell-Garnett mixing formulae (see Appendix A). For the case of neutral inclusions ($\kappa_1 = \kappa_0$ and $\sigma_2 = \sigma_0$) with $\kappa_2 \ll \kappa_0$ and $\sigma_1 \ll \sigma_0$, Fig. 7(a) shows the anisotropy-ratio $\kappa_{\rho,eff}/\kappa_{\phi,eff}$ or $\sigma_{\rho,eff}/\sigma_{\phi,eff}$ that can be attained from Eqs. (8)-(11) by letting the depolarization factors and filling fractions vary within the above mentioned ranges. We emphasize that the neutral-inclusion assumption decouples the syntheses in the thermal and electrical domains, so that the results in Fig. 7(a) are valid for either the thermal parameters ($\kappa_{\rho,eff}/\kappa_{\phi,eff}$, assuming $\Gamma_{2\phi}$ on the abscissa) or the electrical parameters ($\sigma_{\rho,eff}/\sigma_{\phi,eff}$, assuming $\Gamma_{1\phi}$ on the abscissa). Besides confirming the anticipated trends (with the extreme values observed at the extrema of the allowed range for $\Gamma_\phi$, and improving for increasing values of the filling fractions), these results also quantify the attainable anisotropy ratios to moderate values ranging from $\sim 0.25$ to $\sim 2.5$. In order to translate these bounds to the space of coordinate-transformations $F(\rho)$ that can be implemented, we note from Eqs. (15) that the anisotropy ratios directly affect the function $F(\rho)/[\rho\dot{F}(\rho)]$. It makes therefore sense to represent these bounds in the two-dimensional space $\dot{F}(\rho)$ vs. $F(\rho)/\rho$ illustrated in Fig. 7(b). In such space, a given anisotropy ratio corresponds to a straight line passing through the origin, with the slope decreasing with increasing values of the anisotropy-ratio $\kappa_{\rho,eff}/\kappa_{\phi,eff}$ (or $\sigma_{\rho,eff}/\sigma_{\phi,eff}$). In particular, the (dashed) bisector represents the identity transformation $F(\rho) = \rho$. Thus, for fixed filling fractions ($f_1 = f_2 = 0.2$), by drawing the lines corresponding to the maximum and minimum anisotropy ratios attainable [extracted from Fig. 7(a)], we obtain an angular sector [cyan shaded in Fig. 7(b)] which contains all the possible combinations between $\dot{F}(\rho)$ and $F(\rho)/\rho$ that can be implemented within the assumed parameter constraints. To give an idea, also shown in Fig. 7(b) are the curves pertaining to the ideal concentrator (red) and cloak (blue) coordinate-transformations in Fig. 3(a). It can be observed that only a portion of these curves actually falls within the allowed angular sector. The inset shows a magnified view of these portions, with the markers corresponding to the discretized samples in Fig. 3(b), which were purposely chosen so as to fall within the allowed region.

The above analysis provides useful indications for synthesizing more general functionalities, different from the concentrator and cloak in the chosen example.

## V. CONCLUSIONS AND PERSPECTIVES

The characteristics that we have illustrated in this study are a vivid example of artificial structures collectively transcending their natural limitations, and doing so in multiple physical domains independently and simultaneously. The integration of this concept in advanced materials such as ceramics, polymers, biomaterials, and thin films can span multiple orders of magnitude in material scales (from atomic and molecular level to macroscale composites) and may be leveraged to design indiscrete structures from the ground up while bringing about new dimensionalities. Hybrid metamaterials where functional substances are embedded in bigger artificial hetero-structures to induce another level of functionalities, for instance, can now be taken to multiple physical domains to bring more sophistication to material properties.

The transformation-multiphysics framework presented here may be extended and applied to a multitude of electrical, magnetic, acoustic, and thermal systems in various combinations, in both static equilibrium and dynamic non-equilibrium states. In the case of designing a material to manipulate electrical and thermal currents independently, applications may range from multi-functional electronic components to properly engineered thermoelectric materials that affect the figure of merit in ways unexplored in the past. We are currently exploring these possibilities theoretically as well as experimentally.

Just as the transformation-optics paradigm has opened a new door to artificial materials with unconventional attributes, material engineering based on simultaneous coordinate transformations in multiple physical domains may lead to various new possibilities for material characteristics that never existed in the past.

**ACKNOWLEDGMENTS**

S. S. and Y. S. acknowledge support from the Rowland Institute at Harvard University.

**APPENDIX A: DETAILS ON THE EFFECTIVE-MEDIUM FORMULATION**

In our approach, the effective thermal and electrical constitutive parameters of a multiphase mixture composed of $N$ types of inclusions embedded in a host material are modeled via simple Maxwell-Garnett mixing formulae [35], With specific reference to the cylindrical geometry of interest for our study, with inclusions aligned along the local $\rho$ (and $\phi$) direction [cf. Fig. 4(d)], we obtain for the relevant constitutive-tensor components [35]

$$\frac{\kappa_{\rho,eff}}{\kappa_0} = 1 + \frac{\sum_{n=1}^{N} \frac{f_n(\kappa_n - \kappa_0)}{\kappa_0 + \Gamma_{n\rho}(\kappa_n - \kappa_0)}}{1 - \sum_{n=1}^{N} \frac{f_n \Gamma_{n\rho}(\kappa_n - \kappa_0)}{\kappa_0 + \Gamma_{n\rho}(\kappa_n - \kappa_0)}}, \quad \frac{\kappa_{\phi,eff}}{\kappa_0} = 1 + \frac{\sum_{n=1}^{N} \frac{f_n(\kappa_n - \kappa_0)}{\kappa_0 + \Gamma_{n\phi}(\kappa_n - \kappa_0)}}{1 - \sum_{n=1}^{N} \frac{f_n \Gamma_{n\phi}(\kappa_n - \kappa_0)}{\kappa_0 + \Gamma_{n\phi}(\kappa_n - \kappa_0)}}, \quad (A1)$$

$$\frac{\sigma_{\rho,eff}}{\sigma_0} = 1 + \frac{\sum_{n=1}^{N} \frac{f_n(\sigma_n - \sigma_0)}{\sigma_0 + \Gamma_{n\rho}(\sigma_n - \sigma_0)}}{1 - \sum_{n=1}^{N} \frac{f_n \Gamma_{n\rho}(\sigma_n - \sigma_0)}{\sigma_0 + \Gamma_{n\rho}(\sigma_n - \sigma_0)}}, \quad \frac{\sigma_{\phi,eff}}{\sigma_0} = 1 + \frac{\sum_{n=1}^{N} \frac{f_n(\sigma_n - \sigma_0)}{\sigma_0 + \Gamma_{n\phi}(\sigma_n - \sigma_0)}}{1 - \sum_{n=1}^{N} \frac{f_n \Gamma_{n\phi}(\sigma_n - \sigma_0)}{\sigma_0 + \Gamma_{n\phi}(\sigma_n - \sigma_0)}}, \quad (A2)$$

with all parameters already defined in Sec. II-B. In particular, $\Gamma_{n\rho}$ and $\Gamma_{n\phi} = 1 - \Gamma_{n\rho}$ denote the components of the (diagonal, in the cylindrical reference system) depolarization tensor $\vec{\vec{\Gamma}}_n$ pertaining to the generic *n*-th inclusion. From Eqs. (A1) and (A2), it becomes now evident the nonlinear character of the general synthesis problem in Eqs. (6), as well as the coupling between the thermal and electrical domains, as highlighted in Sec. II-B.

The effective-medium model above relies on the calculation of the static polarizabilities of the inclusions, assuming that each inclusion is embedded in an *infinite* host medium. While this may be an acceptable assumption for *sparse* mixtures, it becomes inaccurate for *densely packed* inclusions, for which the medium effectively "seen" outside the generic inclusion is different from the host medium. In this latter scenario, more refined models can be applied such as, e.g., the Polder – van Santen mixing formulae, which approximate the "apparent" medium outside the inclusions as something in between the host medium and the effective medium [35]. However, this yields implicit equations that need to be solved numerically. Acknowledging these limitations, in our approach, we restrict the structural parameters of the mixtures so as to avoid the dense-packing conditions, and remain within the range of applicability of the Maxwell-Garnett mixing formulae in Eqs. (A1) and (A2). As anticipated in Sec. II-B (and better quantified in Sec. IV-B), this inherently limits the attainable anisotropy ratios to *moderate* values.

The three-phase mixture considered in Sec. II-B is the simplest reduction ($N=2$) of Eqs. (A1) and (A2) that still allows the joint and independent synthesis of different functionalities in the thermal and electrical domains. To better understand this aspect, we first consider the simpler two-phase mixture, i.e., with only one type of inclusions embedded in a host medium. By particularizing Eqs. (A1) to this case ($N=1$), we obtain for the thermal parameters

$$\frac{\kappa_{\rho,\text{eff}}}{\kappa_0} = 1 + \frac{\dfrac{f_1(\kappa_1-\kappa_0)}{\kappa_0+\Gamma_{1\rho}(\kappa_1-\kappa_0)}}{1-\dfrac{f_1\Gamma_{1\rho}(\kappa_1-\kappa_0)}{\kappa_0+\Gamma_{1\rho}(\kappa_1-\kappa_0)}} = \frac{\bar{\kappa}_1 - f_0\Gamma_{1\phi}(\bar{\kappa}_1-1)}{1+f_0\Gamma_{1\rho}(\bar{\kappa}_1-1)}, \tag{A3}$$

$$\frac{\kappa_{\phi,\text{eff}}}{\kappa_0} = 1 + \frac{\dfrac{f_1(\kappa_1-\kappa_0)}{\kappa_0+\Gamma_{1\phi}(\kappa_1-\kappa_0)}}{1-\dfrac{f_1\Gamma_{1\phi}(\kappa_1-\kappa_0)}{\kappa_0+\Gamma_{1\phi}(\kappa_1-\kappa_0)}} = \frac{\bar{\kappa}_1 - f_0\Gamma_{1\rho}(\bar{\kappa}_1-1)}{1+f_0\Gamma_{1\phi}(\bar{\kappa}_1-1)}, \tag{A4}$$

with $\bar{\kappa}_1 = \kappa_1/\kappa_0$, and the second equalities following from the consistency conditions $f_0+f_1=1$ and $\Gamma_{1\rho}+\Gamma_{1\phi}=1$. Similarly, for the electrical parameters, we obtain from Eqs. (A2)

$$\frac{\sigma_{\rho,\text{eff}}}{\sigma_0} = 1 + \frac{\dfrac{f_1(\sigma_1-\sigma_0)}{\sigma_0+\Gamma_{1\rho}(\sigma_1-\sigma_0)}}{1-\dfrac{f_1\Gamma_{1\rho}(\sigma_1-\sigma_0)}{\sigma_0+\Gamma_{1\rho}(\sigma_1-\sigma_0)}} = \frac{\bar{\sigma}_1 - f_0\Gamma_{1\phi}(\bar{\sigma}_1-1)}{1+f_0\Gamma_{1\rho}(\bar{\sigma}_1-1)}, \tag{A5}$$

$$\frac{\sigma_{\phi,\text{eff}}}{\sigma_0} = 1 + \frac{\dfrac{f_1(\sigma_1-\sigma_0)}{\sigma_0+\Gamma_{1\phi}(\sigma_1-\sigma_0)}}{1-\dfrac{f_1\Gamma_{1\phi}(\sigma_1-\sigma_0)}{\sigma_0+\Gamma_{1\phi}(\sigma_1-\sigma_0)}} = \frac{\bar{\sigma}_1 - f_0\Gamma_{1\rho}(\bar{\sigma}_1-1)}{1+f_0\Gamma_{1\phi}(\bar{\sigma}_1-1)}, \tag{A6}$$

with $\bar{\sigma}_1 = \sigma_1/\sigma_0$. We note that, for the extreme values $\Gamma_{1\phi}=0$ and $\Gamma_{1\phi}=1$, Eqs. (A3)-(A6) reduce to the well-known expressions pertaining to radial and angular multilayers [35], respectively, which have been widely utilized to design coordinate-transformation-inspired metamaterial structures implementing single functionalities (e.g., cloak, concentrator) in the thermal, electrical, or magnetic domains [17-24,26-28]. However, it can be verified that the mixing formulae in Eqs. (A3)-(A6) do not provide enough degrees of freedom to design different anisotropic behaviors in the thermal and electrical domains. For instance, assuming that the parameters $\bar{\kappa}_1$, $f_0$ and $\Gamma_{1\phi}$ in Eqs. (A3) and (A4) are chosen so as to guarantee that $\kappa_{\rho,eff}/\kappa_{\phi,eff} > 1$, it will be then impossible to achieve an anisotropy ratio $\sigma_{\rho,eff}/\sigma_{\phi,eff} < 1$ from Eqs. (A5) and (A6), for *any* choice (subject to the passivity condition) of the material parameter $\bar{\sigma}_1$. This can be verified in a rather straightforward fashion for the limit (multilayer) cases $\Gamma_{1\phi} = 0$ or $\Gamma_{1\phi} = 1$, and in a more cumbersome fashion (we relied on the "Reduce" symbolic-algebra tool of Mathematica™ [43]) for general values of $\Gamma_{1\phi}$. Clearly, it represents a significant curtail of the capabilities to independently manipulate the phenomena in the two physical domains. For instance, it is clear from Fig. 3(b) that the joint synthesis of a thermal concentrator (which requires $\kappa_\rho/\kappa_\phi > 1$) and an electrical cloak (which requires $\sigma_\rho/\sigma_\phi < 1$) would not be possible with this type of mixtures. It is worth highlighting that these constraints may be relaxed in the presence of *negative-conductivity* material constituents. While these materials are not available in Nature, they may be in turn synthesized as metamaterials. For instance, Fang *et. al* [19] recently demonstrated experimentally the possibility of synthesizing an artificial material exhibiting negative electrical conductivity, by means of active devices together with resistor networks.

An easier way to overcome the above limitations, while maintaining the passivity requirements, entails considering a three-phase mixture, featuring two types of inclusions embedded in a host medium. For this scenario ($N = 2$), we now obtain from Eqs. (A1) and (A2):

$$\frac{\kappa_{\rho,eff}}{\kappa_0} = 1 + \frac{\dfrac{f_1(\kappa_1 - \kappa_0)}{\kappa_0 + \Gamma_{1\rho}(\kappa_1 - \kappa_0)} + \dfrac{f_2(\kappa_2 - \kappa_0)}{\kappa_0 + \Gamma_{2\rho}(\kappa_2 - \kappa_0)}}{1 - \dfrac{f_1 \Gamma_{1\rho}(\kappa_1 - \kappa_0)}{\kappa_0 + \Gamma_{1\rho}(\kappa_1 - \kappa_0)} - \dfrac{f_2 \Gamma_{2\rho}(\kappa_2 - \kappa_0)}{\kappa_0 + \Gamma_{2\rho}(\kappa_2 - \kappa_0)}}, \tag{A7}$$

$$\frac{\kappa_{\phi,eff}}{\kappa_0} = 1 + \frac{\dfrac{f_1(\kappa_1 - \kappa_0)}{\kappa_0 + \Gamma_{1\phi}(\kappa_1 - \kappa_0)} + \dfrac{f_2(\kappa_2 - \kappa_0)}{\kappa_0 + \Gamma_{2\phi}(\kappa_2 - \kappa_0)}}{1 - \dfrac{f_1 \Gamma_{1\phi}(\kappa_1 - \kappa_0)}{\kappa_0 + \Gamma_{1\phi}(\kappa_1 - \kappa_0)} - \dfrac{f_2 \Gamma_{2\phi}(\kappa_2 - \kappa_0)}{\kappa_0 + \Gamma_{2\phi}(\kappa_2 - \kappa_0)}}, \tag{A8}$$

$$\frac{\sigma_{\rho,eff}}{\sigma_0} = 1 + \frac{\dfrac{f_1(\sigma_1-\sigma_0)}{\sigma_0+\Gamma_{1\rho}(\sigma_1-\sigma_0)} + \dfrac{f_2(\sigma_2-\sigma_0)}{\sigma_0+\Gamma_{2\rho}(\sigma_2-\sigma_0)}}{1 - \dfrac{f_1\Gamma_{1\rho}(\sigma_1-\sigma_0)}{\sigma_0+\Gamma_{1\rho}(\sigma_1-\sigma_0)} - \dfrac{f_2\Gamma_{2\rho}(\sigma_2-\sigma_0)}{\sigma_0+\Gamma_{2\rho}(\sigma_2-\sigma_0)}}, \tag{A9}$$

$$\frac{\sigma_{\phi,eff}}{\sigma_0} = 1 + \frac{\dfrac{f_1(\sigma_1-\sigma_0)}{\sigma_0+\Gamma_{1\phi}(\sigma_1-\sigma_0)} + \dfrac{f_2(\sigma_2-\sigma_0)}{\sigma_0+\Gamma_{2\phi}(\sigma_2-\sigma_0)}}{1 - \dfrac{f_1\Gamma_{1\phi}(\sigma_1-\sigma_0)}{\sigma_0+\Gamma_{1\phi}(\sigma_1-\sigma_0)} - \dfrac{f_2\Gamma_{2\phi}(\sigma_2-\sigma_0)}{\sigma_0+\Gamma_{2\phi}(\sigma_2-\sigma_0)}}. \tag{A10}$$

While it is now possible, in principle, to achieve different anisotropy ratios in the thermal and electrical domains, the general synthesis in Eqs. (6) remains a formidable task, which, in the general case, can be addressed in a weak fashion, i.e., by minimizing a suitable cost function parameterizing the mismatch of the effective and nominal parameters, and possibly including regularization terms.

The synthesis problem is significantly simplified if we consider "neutral" inclusions [36,37], i.e., inclusions that are matched with the background medium in one physical domain. Assuming, for instance, $\kappa_1 = \kappa_0$ (i.e., thermally-neutral type-1 inclusions) and $\sigma_2 = \sigma_0$ (i.e., electrically-neutral type-2 inclusions), Eqs. (A7)-(A10) reduce to the forms in Eqs. (8)-(11), which effectively decouple the synthesis problem. By substituting Eqs. (8)-(11) in Eqs. (6) (and letting $\kappa_{\rho,nom}$, $\kappa_{\phi,nom}$, $\sigma_{\rho,nom}$, and $\sigma_{\phi,nom}$ the relevant components of the nominal parameters to synthesize), we obtain four equations that can be solved analytically in closed form. More specifically, the depolarization factors can be found as solutions of second-degree equations, viz.,

$$\Gamma_{2\phi} = \frac{(1-\bar{\kappa}_2)\left[1-\bar{\kappa}_{nom}(1-2f_2)\right] \pm \sqrt{(1+\bar{\kappa}_2)^2(1-\bar{\kappa}_{nom})^2 + 4f_2^2\bar{\kappa}_{nom}(1-\bar{\kappa}_2)^2}}{2(1-\bar{\kappa}_2)(1-f_2)(1-\bar{\kappa}_{nom})}, \tag{A11}$$

$$\Gamma_{1\phi} = \frac{(1-\bar{\sigma}_1)\left[1-\bar{\sigma}_{nom}(1-2f_1)\right] \pm \sqrt{(1+\bar{\sigma}_1)^2(1-\bar{\sigma}_{nom})^2 + 4f_1^2\bar{\sigma}_{nom}(1-\bar{\sigma}_1)^2}}{2(1-\bar{\sigma}_1)(1-f_1)(1-\bar{\sigma}_{nom})}, \tag{A12}$$

from which it is rather straightforward to find the constitutive parameters

$$\kappa_0 = \kappa_1 = \kappa_{\phi,nom} \frac{1+\Gamma_{2\phi}(1-f_2)(\bar{\kappa}_2-1)}{1+\Gamma_{2\phi}(1-f_2)(\bar{\kappa}_2-1)+f_2(\bar{\kappa}_2-1)}$$
$$= \frac{2\kappa_{\rho,nom}\left[1+\bar{\kappa}_2+(1-\bar{\kappa}_2)f_2\right]}{(1+\bar{\kappa}_2)(1+\bar{\kappa}_{nom})\pm\sqrt{(1+\bar{\kappa}_2)^2(1-\bar{\kappa}_{nom})^2+4f_2^2\bar{\kappa}_{nom}(1-\bar{\kappa}_2)^2}},$$
(A13)

$$\sigma_0 = \sigma_2 = \sigma_{\phi,nom} \frac{1+\Gamma_{1\phi}(1-f_1)(\bar{\sigma}_1-1)}{1+\Gamma_{1\phi}(1-f_1)(\bar{\sigma}_1-1)+f_1(\bar{\sigma}_1-1)}$$
$$= \frac{2\sigma_{\rho,nom}\left[1+\bar{\sigma}_1+(1-\bar{\sigma}_1)f_1\right]}{(1+\bar{\sigma}_1)(1+\bar{\sigma}_{nom})\pm\sqrt{(1+\bar{\sigma}_1)^2(1-\bar{\sigma}_{nom})^2+4f_1^2\bar{\sigma}_{nom}(1-\bar{\sigma}_1)^2}}.$$
(A14)

In Eqs. (A11)-(A14), $\bar{\kappa}_2 = \kappa_2/\kappa_0$, $\bar{\sigma}_1 = \sigma_1/\sigma_0$, $\bar{\kappa}_{nom} = \kappa_{\rho,nom}/\kappa_{\phi,nom}$, $\bar{\sigma}_{nom} = \sigma_{\rho,nom}/\sigma_{\phi,nom}$, and the $\pm$ sign is consistently chosen so as to ensure the passivity ($\kappa_0 \geq 0$, $\sigma_0 \geq 0$) and model-consistency ($0 < \Gamma_{1\phi} < 1$, $0 < \Gamma_{2\phi} < 1$) conditions. We note that the above solutions contain as free parameters the normalized conductivities $\bar{\kappa}_2$ and $\bar{\sigma}_2$ (subject to the passivity conditions $\bar{\kappa}_2 \geq 0$, $\bar{\sigma}_2 \geq 0$) as well as the fractions $f_1$ and $f_2$ (subject to $0 < f_1 < 1$, $0 < f_2 < 1$, and $f_0 + f_1 + f_2 = 1$). The further simplified expressions in Eqs. (12) and (13) immediately follow by setting $\bar{\kappa}_2 = \bar{\sigma}_1 = 0$ in Eqs. (A11)-(A14). Similarly, alternative simplified expressions can be derived in the opposite asymptotic limit $\bar{\kappa}_2 \gg 1$ and $\bar{\sigma}_2 \gg 1$, viz.

$$\kappa_1 = \kappa_0 = \kappa_{\phi,nom}\frac{\Gamma_{2\phi}(1-f_2)}{\Gamma_{2\phi}(1-f_2)+f_2},$$
(A15)

$$\sigma_2 = \sigma_0 = \sigma_{\phi,nom}\frac{\Gamma_{1\phi}(1-f_1)}{\Gamma_{1\phi}(1-f_1)+f_1},$$
(A16)

with $\Gamma_{1\phi}$ and $\Gamma_{2\phi}$ still given by Eqs. (12) and (13). Clearly, mixed limits, such as $\bar{\kappa}_2 \ll 1$ and $\bar{\sigma}_2 \gg 1$, or $\bar{\kappa}_2 \gg 1$ and $\bar{\sigma}_2 \ll 1$, may also be derived.

### APPENDIX B: DETAILS ON THE COORDINATE TRANSFORMATIONS

As mentioned in Sec. III-A, the thermal-concentration and electrical-cloak effects are essentially established by the boundary values of the mapping functions in Eqs. (16), whereas the function behaviors in between are only partially constrained by the continuity requirement (in order to avoid additional boundary conditions) as well as by the nonnegative character of their logarithmic derivatives [$F_t/\dot{F}_t \geq 0$, $F_e/\dot{F}_e \geq 0$, in order to guarantee passivity, cf. Eqs. (15)]. The cloak transformation utilized in our study [blue curve in Fig. 3(a)] belongs to the general class of algebraic transformations

$$F_e(\rho) = R_2\left(\frac{\rho - R_1}{R_2 - R_1}\right)^\gamma, \quad \gamma > 0, \tag{B1}$$

which satisfy the required boundary conditions (16), and map a cylinder of radius $R_2$ in the auxiliary space onto an annular cylinder of radii $R_1$ and $R_2$ in the transformed space, thereby creating a "hole" of radius $R_1$ that admits no image in the auxiliary space. For $\gamma = 1$, Eq. (B1) reduces to the standard (linear) cloak transformation introduced by Pendry *et al.* [2]. Here, we consider instead

$$\gamma = 1 - \frac{R_1}{R_2}, \tag{B2}$$

which yields

$$\dot{F}_e(R_2) = \frac{F_e(R_2)}{R_2} = 1. \tag{B3}$$

Recalling Eqs. (15), this ensures that

$$\sigma_\rho(R_2) = \sigma_\phi(R_2) = \sigma', \tag{B4}$$

i.e, that the arising transformation medium tends (for $\rho \to R_2$) to an isotropic material matched with the background medium. As mentioned in Sec. III, while not strictly necessary, this assumption simplifies the inclusion-based implementation.

For the concentrator transformation, we exploit the degrees of freedom in the choice of the mapping function, by enforcing the condition in Eq. (17). Looking at Eqs. (12) and (13), it can be observed that with this assumption (i.e., $\bar{\kappa}_{nom} = 1/\bar{\sigma}_{nom}$), together with $f_1 = f_2$, we obtain $\Gamma_{2\phi} = 1 - \Gamma_{1\phi} = \Gamma_{1\rho}$, which means that the two types of inclusions have identical shape (just rotated of 90°). Once again, while not strictly necessary, the above assumption may facilitate the inclusion-based implementation, allowing more efficient packing strategies. Recalling Eqs. (6), the condition in Eq. (17) yields the differential equation

$$\frac{\rho^2 \dot{F}_e(\rho)}{F_e(\rho)} \dot{F}_t(\rho) - F_t(\rho) = 0, \tag{B5}$$

with the boundary condition $F_t(R_2) = R_2$. By substituting Eq. (B1) in Eq. (B5), we then obtain

$$\frac{\gamma \rho^2}{(\rho - R_1)} \dot{F}_t(\rho) - F_t(\rho) = 0, \tag{B6}$$

which admits the simple analytical solution

$$F_t(\rho) = R_2 \left(\frac{\rho}{R_2}\right)^{\frac{1}{\gamma}} \exp\left[\frac{R_1(R_2 - \rho)}{\gamma R_2 \rho}\right] \tag{B7}$$

considered in our study [red curve in Fig. 3(a)]. From Eq. (B7), we can easily calculate the concentration factor,

$$c = \frac{F_t(R_1)}{R_1} = e\left(\frac{R_1}{R_2}\right)^{\frac{R_1}{R_2 - R_1}} > 1, \tag{B8}$$

thereby verifying the concentrator functionality.

# REFERENCES

[1] U. Leonhardt, *Optical Conformal Mapping*, Science **312**, 1777 (2006).

[2] J. B. Pendry, D. Schurig, and D. R. Smith, *Controlling Electromagnetic Fields*, Science **312**, 1780 (2006).

[3] D. Schurig, J. J. Mock, B. J. Justice, S. A. Cummer, J. B. Pendry, A. F. Starr, and D. R. Smith, *Metamaterial Electromagnetic Cloak at Microwave Frequencies*, Science **314**, 977 (2006).

[4] H. Chen, C. T. Chan, and P. Sheng, *Transformation Optics and Metamaterials*, Nature Mater. **9**, 387 (2010).

[5] D. H. Werner and D. H. Kwon, Eds., *Transformation Electromagnetics and Metamaterials* (Springer, Berlin-Heildelberg, 2013).

[6] S. A. Cummer and D. Schurig, *One Path to Acoustic Cloaking*, New J. Phys. **9**, 45 (2007).

[7] H. Chen and C. T. Chan, *Acoustic Cloaking in Three Dimensions Using Acoustic Metamaterials*, Appl. Phys. Lett. **91,** 183518 (2007).

[8] A. N. Norris, *Acoustic Cloaking Theory*, Proc. R. Soc. A **464**, 2411 (2008).

[9] H. Chen and C. T. Chan, *Acoustic Cloaking and Transformation Acoustics*, J. Phys. D: Appl. Phys. **43**, 113001 (2010).

[10] R. V. Craster and S. Guenneau, Eds., *Acoustic Metamaterials* (Springer Netherlands, Dordrecht, 2013).

[11] G. W. Milton, M. Briane, and J. R. Willis, *On Cloaking for Elasticity and Physical Equations with a Transformation Invariant Form*, New J. Phys. **8**, 248 (2006).

[12] A. N. Norris and A. L. Shuvalov, *Elastic Cloaking Theory*, Wave Motion **48**, 525 (2011).

[13] M. Brun, S. Guenneau, and A. B. Movchan, *Achieving Control of In-Plane Elastic Waves*, Appl. Phys. Lett. **94**, 061903 (2009).

[14] M. Farhat, S. Guenneau, and S. Enoch, *Ultrabroadband Elastic Cloaking in Thin Plates*, Phys. Rev. Lett. **103**, 024301 (2009).

[15] A. N. Norris and W. J. Parnell, *Hyperelastic Cloaking Theory: Transformation Elasticity with Pre-Stressed Solids*, Proc. R. Soc. A **468**, 2881 (2012).

[16] N. Stenger, W. Wilhelm, and M. Wegener, *Experiments on Elastic Cloaking in Thin Plates*, Phys. Rev. Lett. **108**, 014301 (2012).

[17] A. Greenleaf, M. Lassas, and G. Uhlmann, *Anisotropic Conductivities that Cannot be*
22

| | Radii (cm) | | Host | Inclusions | | | | Fractions |
|---|---|---|---|---|---|---|---|---|
| | | | | Type 1 ($\kappa_1=\kappa_0, \sigma_1=0$) | | Type 2 ($\kappa_2=0, \sigma_2=\sigma_0$) | | |
| Layer | $R_{in}$ | $R_{out}$ | $\kappa_0/\kappa' = \sigma_0/\sigma'$ | $\Gamma_{1\phi}$ | $A_{1\rho}/A_{1\phi}$ | $\Gamma_{2\phi}$ | $A_{2\rho}/A_{2\phi}$ | $f_1 = f_2$ |
| 1 | 2 | 4.86 | 1.875 | 0.121 | 0.138 | 0.879 | 7.255 | 0.18 |
| 2 | 4.86 | 5.60 | 1.630 | 0.147 | 0.172 | 0.853 | 5.813 | 0.15 |
| 3 | 5.60 | 6.36 | 1.416 | 0.144 | 0.168 | 0.856 | 5.969 | 0.10 |
| 4 | 6.36 | 7.14 | 1.307 | 0.158 | 0.188 | 0.842 | 5.311 | 0.08 |
| 5 | 7.14 | 7.94 | 1.221 | 0.166 | 0.199 | 0.834 | 5.035 | 0.06 |
| 6 | 7.94 | 12 | 1.150 | 0.160 | 0.191 | 0.840 | 5.246 | 0.04 |

Table I. Geometrical, structural, and constitutive parameters of the ideal-parameter synthesis, from the piecewise-constant nominal profiles in Fig. 3(b), assuming neutral inclusions ($\kappa_1 = \kappa_0$, $\sigma_2 = \sigma_0$) with $\kappa_2 = 0$ and $\sigma_1 = 0$.

| | Material | $\kappa\,[W/(mK)]$ | $\sigma\,[S/m]$ |
|---|---|---|---|
| **Layers 1 and 2** | PG momentive | 300 | $2 \cdot 10^5$ |
| **Layers 3 and 4** | P-75S 2k | 185 | $1.43 \cdot 10^5$ |
| **Layers 5 and 6, background** | P-55S 2k | 120 | $1.18 \cdot 10^5$ |
| **Type-1 inclusions** | Aluminum nitride | 190 | $10^{-11}$ |
| **Type-2 inclusions** | Silver conductive epoxy | 1.75 | $1.4 \cdot 10^5$ |

Table II. Realistic materials (and corresponding thermal and electrical conductivities [41]) considered for approximating the ideal-parameter synthesis in Table I and Fig. 4.

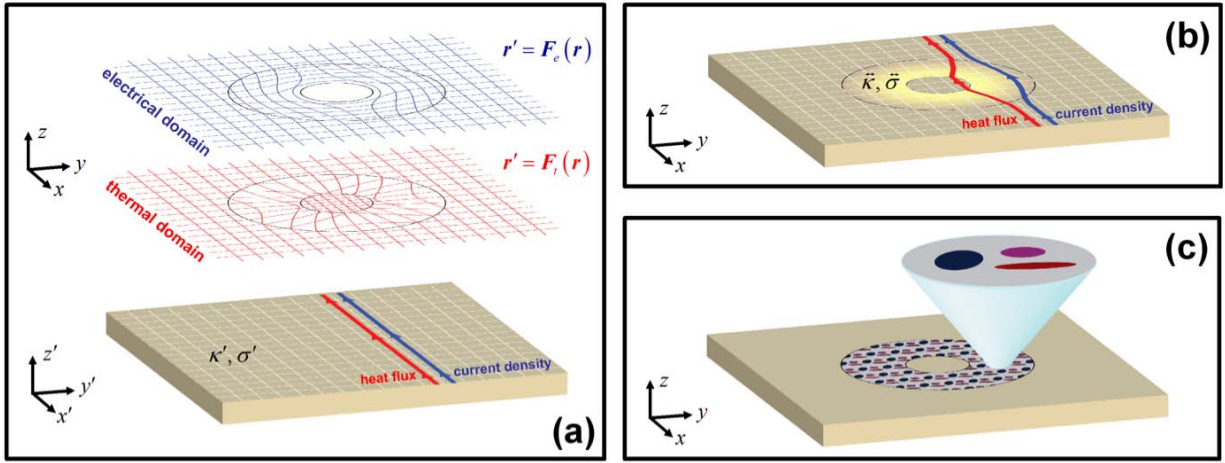

FIG. 1. (a) Auxiliary space filled with an isotropic, homogeneous medium, wherein heat-flux and electrical current-density follow parallel straight paths. Two coordinate transformations are applied which induce different behaviors in the thermal (e.g., concentrator) and electrical (e.g., cloak) domains. (b) Equivalent interpretation, in a flat-metric space filled with a transformation medium [cf. Eqs. (4)]. The heat-flux and current-density paths are distorted in different fashions. (c) Metamaterial-based approximate implementation of the required nominal constitutive parameters via a mixture of inclusions of different shapes and materials (as qualitatively depicted in the magnified details) embedded in a host medium.

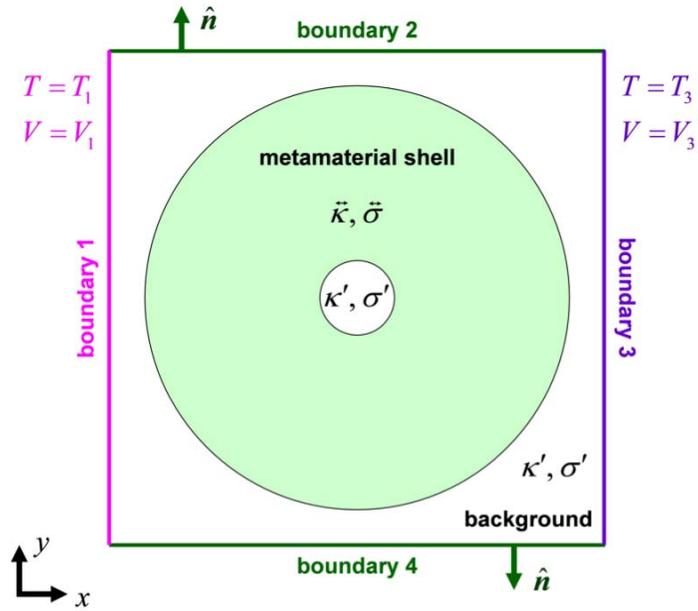

FIG. 2. Schematic of the computational domain considered in the finite-element numerical simulations, consisting of a square of sidelenght $L = 28 cm$ filled with a background material with constitutive parameters $\kappa'$ and $\sigma'$, and a metamaterial annular shell of radii $R_1 = 2 cm$ and $R_2 = 12 cm$. Also indicated are the boundary conditions enforced at the left and right boundaries (1 and 3, respectively), as well as the outward normal unit-vectors involved in the thermal-insuation and electrical-insulation conditions enforced at the boundaries 2 and 4.

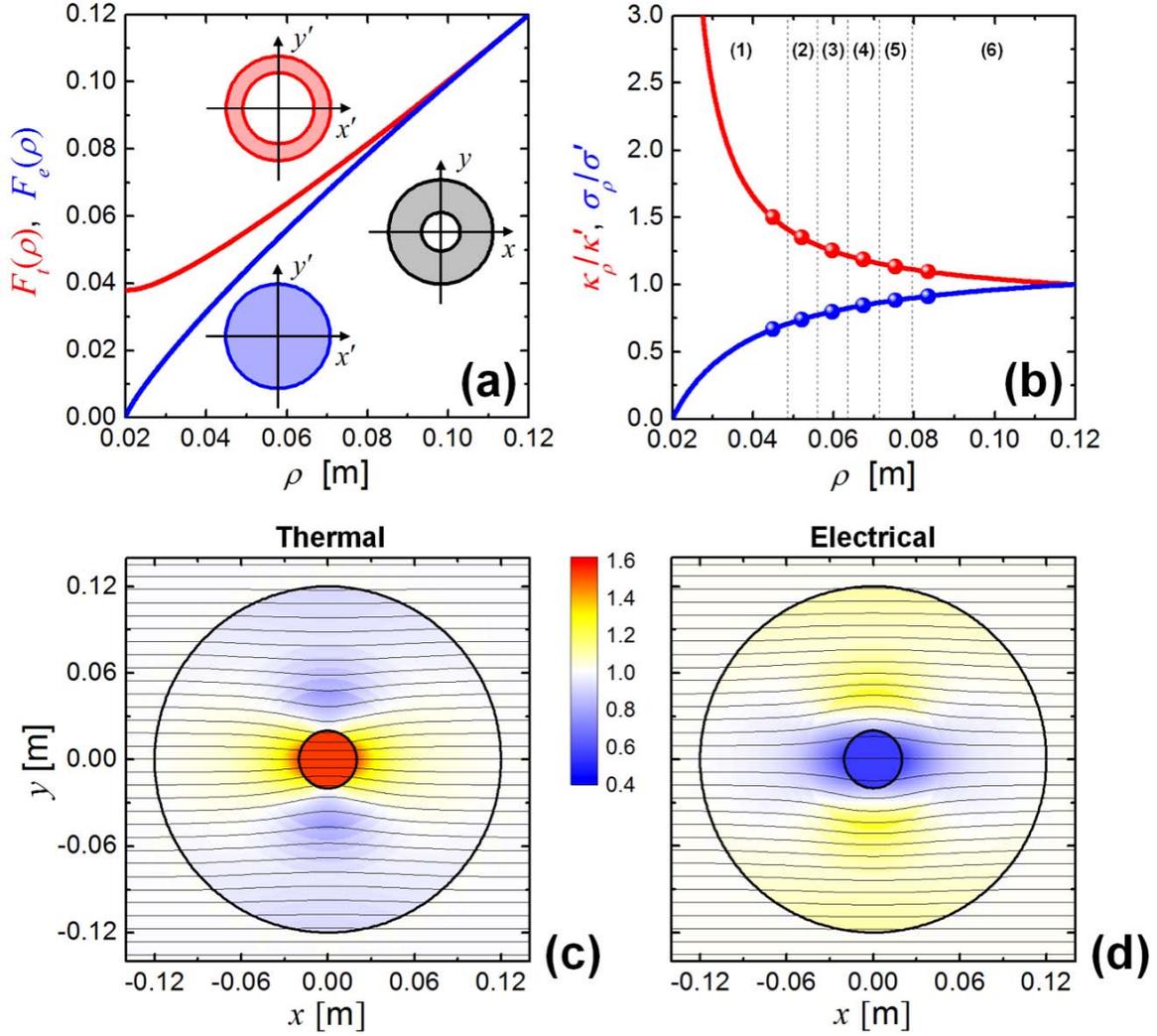

FIG. 3. (a) Radial coordinate transformations implementing the ideal thermal concentrator (red) and electrical cloak (blue), within an annulus of radii $R_1 = 2cm$ and $R_2 = 12cm$. Also shown is a qualitative illustration of the mapping between auxiliary and transformed spaces. (b) Corresponding relevant constitutive-tensor components $\kappa_\rho/\kappa' = \kappa'/\kappa_\phi$ (red) and $\sigma_\rho/\sigma' = \sigma'/\sigma_\phi$ (blue). Outside the annulus $R_1 < \rho < R_2$, the coordinate transformations reduce to the identity, $F_{t,e}(\rho) = \rho$, and the parameters coincide with those in the auxiliary space ($\kappa'$ and $\sigma'$). The vertical dashed lines indicate the six-layer piecewise-constant radial discretization considered, with the markers representing the constant values assumed in each layer. (c), (d) Numerically-computed steady-state total heat-flux and electrical current-density magnitudes, respectively, with the superimposed streamlines indicating the local directions.

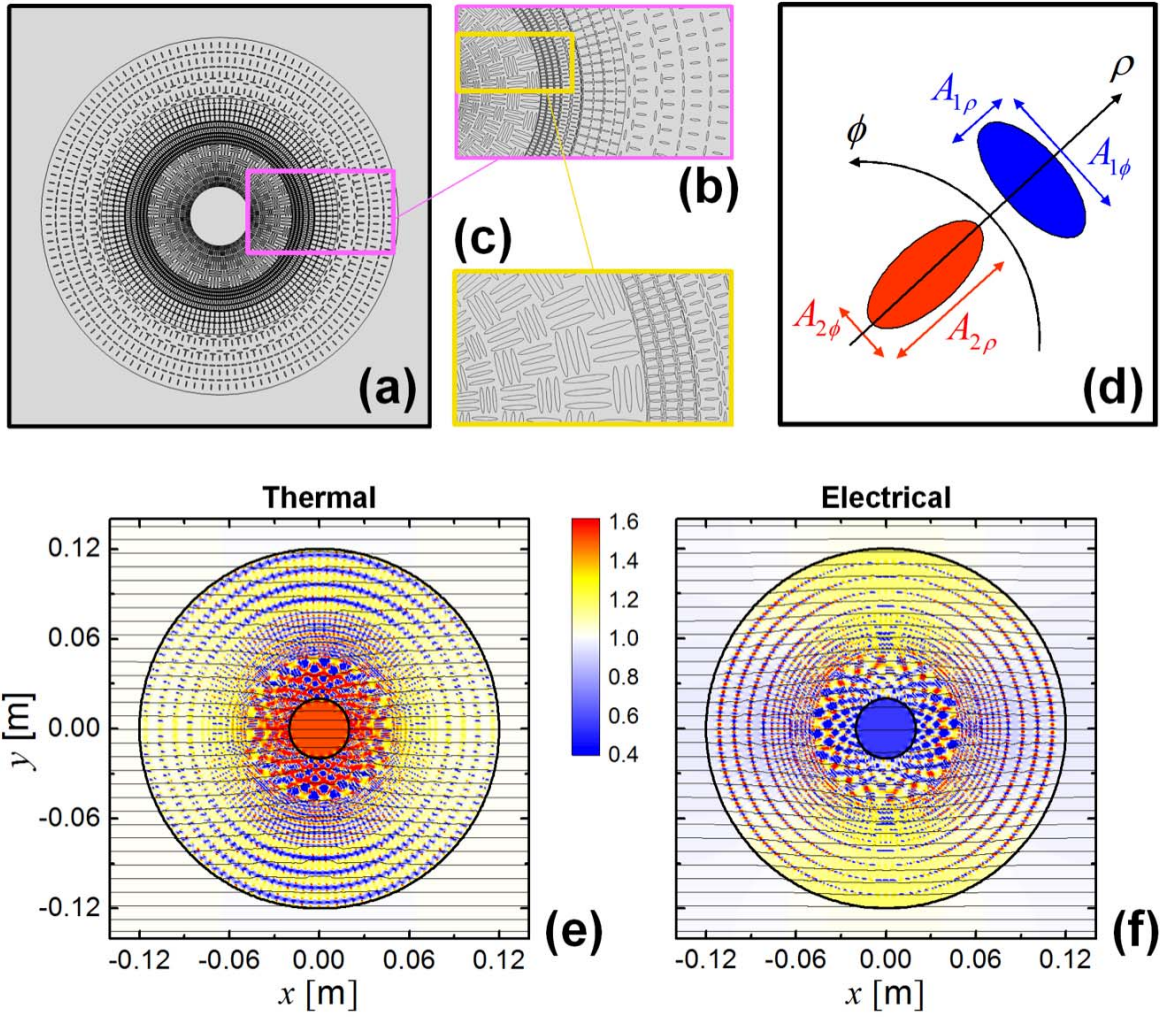

FIG. 4. (a) Geometry of the metamaterial implementation of the piecewise-constant constitutive parameter distributions in Fig. 3(b), based on ideal material constituents (details in Table I). (b), (c) Magnified details of the inclusions. (d) Schematic of the generic type-1 (blue) and type-2 (red) elliptical inclusions. (e), (f) Corresponding thermal and electrical responses, respectively, as in Figs. 3(c) and 3(d), assuming a background medium with $\kappa' = 1\,W/(mK)$ and $\sigma' = 1\,S/m$. Although a moderately larger dynamical range is observed in the metamaterial shell, the same colorscale as in Figs. 3(c) and 3(d) is used, so as to facilitate direct comparison of the quantities in the inner and exterior regions.

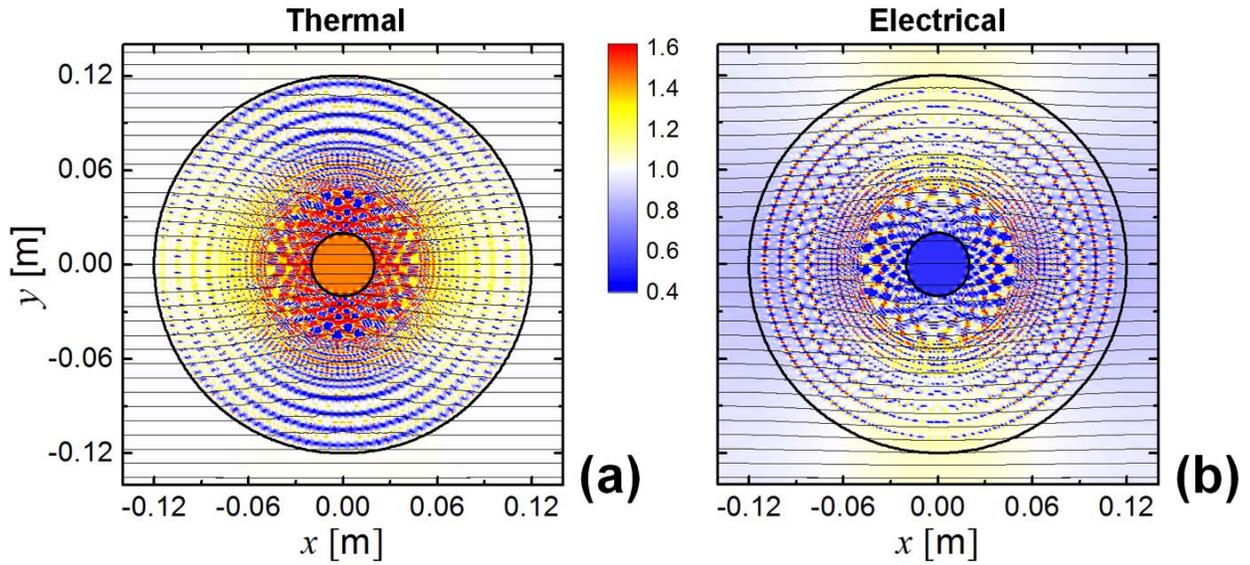

FIG. 5. (a), (b) As in Figs. 4(e) and 4(f), respectively, but assuming the realistic material parameters in Table II.

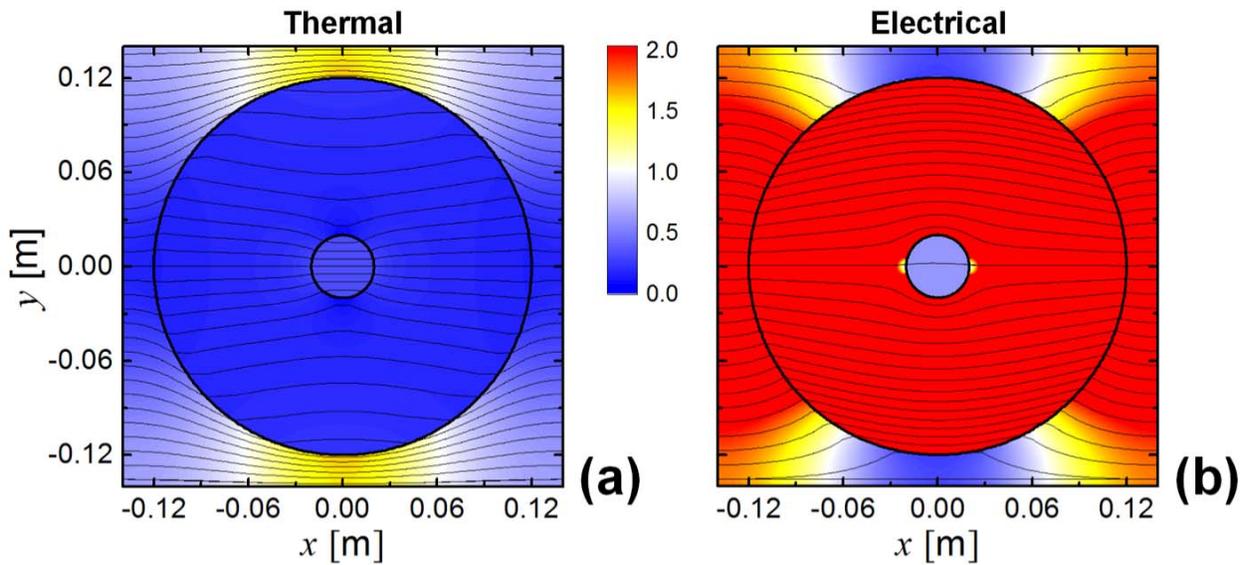

FIG. 6. As in Fig. 5, but for a shell (of same size) made of stainless steel ($\kappa = 16.3\, W/(mK)$, $\sigma = 1.42 \cdot 10^6\, S/m$) [42].

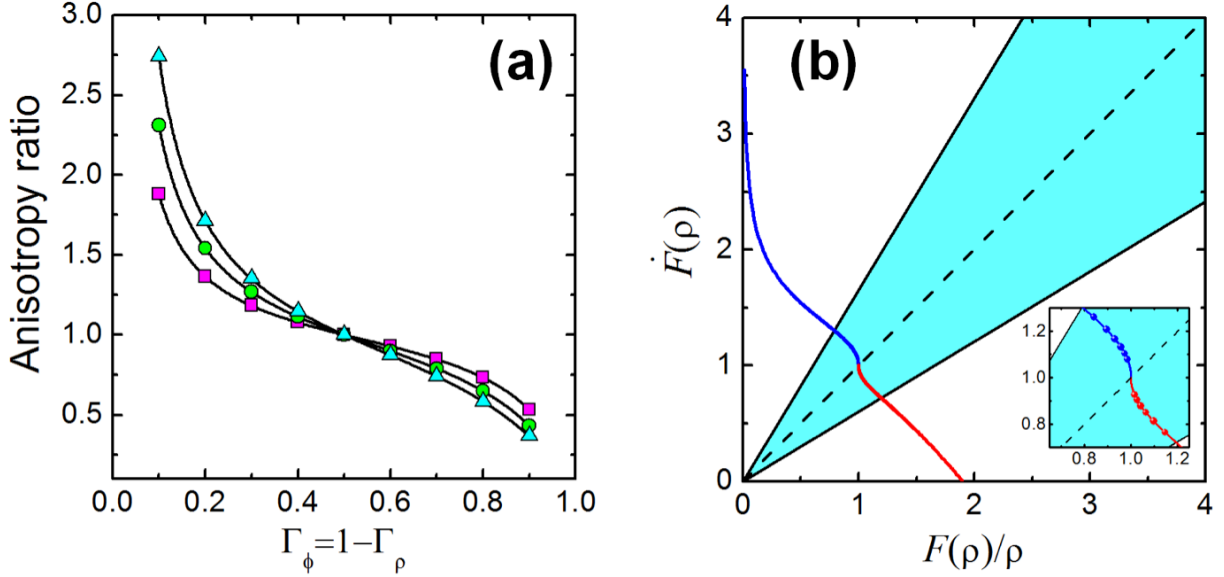

FIG. 7. (a) Anisotropy-ratio $\kappa_{\rho,eff}/\kappa_{\phi,eff}$ or $\sigma_{\rho,eff}/\sigma_{\phi,eff}$ attainable from Eqs. (8)-(11) (for the case of neutral inclusions $\kappa_1 = \kappa_0$ and $\sigma_2 = \sigma_0$, with $\kappa_2 \ll \kappa_0$ and $\sigma_1 \ll \sigma_0$) as a function of the depolarization factor $\Gamma_\phi$, for three representative filling-fraction values $f_1 = f_2 = 0.1, 0.15, 0.2$ (square, circle, and triangle markers, respectively). (b) Representation of the anisotropy-ratio bounds in the space $\dot{F}(\rho)$ vs. $F(\rho)/\rho$, assuming $f_1 = f_2 = 0.2$. The cyan-shaded angular sector represents the allowed region, and the dashed bisector the identity transformation. Also shown are the curves pertaining to the concentrator (red) and cloak (blue) coordinate-transformations in Fig. 3(a). The inset shows a magnified view of the allowed portions of these curves, with the markers corresponding to the discretized samples in Fig. 3(b).